\documentclass[twocolumn,times,final]{elsarticle}
\usepackage{framed,multirow}

\usepackage{amssymb}
\usepackage{latexsym}

\usepackage{url}
\usepackage{xcolor}

\definecolor{newcolor}{rgb}{.8,.349,.1}

\usepackage{tabularx,booktabs}
\usepackage[scr=rsfso]{mathalfa}
\newcommand{\Fusion}[1]{\ensuremath{\mathscr{F}_{#1}}}

\begin{document}

\setcounter{page}{1}

\begin{frontmatter}

\title{Multi-stage intermediate fusion for multimodal learning to classify  non-small cell lung cancer subtypes from CT and PET}

\author[1]{Fatih Aksu}
\ead{fatih.aksu@st.hunimed.eu}
\author[2,3]{Fabrizia Gelardi}
\ead{gelardi.fabrizia@hsr.it}
\author[2,3]{Arturo Chiti}
\ead{chiti.arturo@hsr.it}
\author[4,5]{Paolo Soda\corref{cor1}}
\cortext[cor1]{Corresponding author}
\ead{paolo.soda@umu.se}


\affiliation[1]{organization={Department of Biomedical Sciences, Humanitas University},
                city={Milan}, 
                country={Italy}}

\affiliation[2]{organization={Università Vita-Salute San Raffaele},
                city={Milan}, 
                country={Italy}}

\affiliation[3]{organization={IRCCS Ospedale San Raffaele},
                city={Milan}, 
                country={Italy}}

\affiliation[4]{organization={Unit of Computer Systems and Bioinformatics, Department of Engineering, Università Campus Bio-Medico di Roma},
                city={Rome}, 
                country={Italy}}

\affiliation[5]{organization={Department of Diagnostics and Intervention, Radiation Physics, Umeå University},
                city={Umeå}, 
                country={Sweden}}


\begin{abstract}
Accurate classification of histological subtypes of non-small cell lung cancer (NSCLC) is essential in the era of precision medicine, yet current invasive techniques are not always feasible and may lead to clinical complications. 
This study presents a multi-stage intermediate fusion approach to classify NSCLC subtypes from CT and PET images.
Our method integrates the two modalities at different stages of feature extraction, using voxel-wise fusion to exploit complementary information across varying abstraction levels while preserving spatial correlations. 
We compare our method against unimodal approaches using only CT or PET images to demonstrate the benefits of modality fusion, and further benchmark it against early and late fusion techniques to highlight the advantages of intermediate fusion during feature extraction. 
Additionally, we compare our model with the only existing intermediate fusion method for histological subtype classification using PET/CT images. 
Our results demonstrate that the proposed method outperforms all alternatives across key metrics, with an accuracy and AUC equal to 0.724 and 0.681, respectively. 
This non-invasive approach has the potential to significantly improve diagnostic accuracy, facilitate more informed treatment decisions, and advance personalized care in lung cancer management.
\end{abstract}



\end{frontmatter}


\section{Introduction}
\label{sec:intro}
Lung cancer is a leading cause of cancer-related deaths globally, with estimated age-adjusted incidence and mortality rates of 23.6 and 16.8 per 100,000 people, respectively~\cite{cancerobservatory}. 
Non-small cell lung cancer (NSCLC) accounts for 85\% of primary lung cancers, with adenocarcinoma (ADC) and squamous cell carcinoma (SQC) being the most common subtypes~\cite{travis2015}. 
The two primary histological subtypes not only have different biological characteristics and outcomes, but also different responses to targeted therapies and immunotherapies~\cite{chansky2009international,campbell2016distinct}. 
In the context of early-stage NSCLC, a full histological examination of the primary tumour prior to surgery may be omitted in cases where there is a significant risk of biopsy-related complications and a compelling clinical indication of malignancy based on imaging and clinical findings. 
However, an accurate pathological diagnosis of the primary tumour is essential to determine prognosis and select the most effective therapeutic strategies in patients with clinical stage I-III disease~\cite{postmus2017early}. Traditional methods of identifying these subtypes rely on tissue biopsy and histopathological examination, which are invasive and can carry significant risks for patients~\cite{de2016image}.
Moreover, such techniques often struggle with accuracy due to challenges like small tumor size, the tumor location near the lung's edges or critical structures, and the diverse characteristics of tumors, which can lead to inconsistent results~\cite{han2021histologic}.
These challenges, along with the limitations of current invasive diagnostic methods and the need to avoid such procedures, drive the search for non-invasive approaches to accurately classify NSCLC histological subtypes. 

Positron Emission Tomography combined with Computed Tomography (PET/CT) using the [18F]Fluorodeoxyglucose (FDG) tracer plays a pivotal role in the diagnosis and management of lung cancer, with most patients undergoing this imaging modality prior to the initiation of treatment~\cite{takeuchi2014impact}. 
By integrating the metabolical imaging capabilities of PET with the detailed anatomical imaging from CT, PET/CT offers enhanced precision in tumor staging, significantly improving the detection and localization of loco-regional pathological lymph nodes and distant metastases~\cite{antoch2003non}. 
Moreover, various subtypes of NSCLC exhibit differing characteristics in these radiological images. However, the limited specificity of these features makes it difficult for radiologists to accurately differentiate between NSCLC subtypes~\cite{jiang2014thin}. 

Artificial intelligence (AI)  continues to succeed in a variety of domains, from natural language processing to computer vision.
The capabilities of deep learning (DL) methods have been demonstrated in numerous studies in medical imaging across tasks such as classification, object detection,  segmentation, and image synthesis~\cite{dhar2023challenges,guarrasi2024multimodal}, including applications in cancer research. In this domain, DL techniques have been employed for tasks like denoising~\cite{lei2023ct}, diagnosis~\cite{qin2020fine,aksu2023early}, and prognosis prediction~\cite{zhu2020application,caruso2024deeplearning}, to name a few.
Histological subtype classification is another challenging task in cancer research, which continues to be addressed by researchers using various AI methods and imaging modalities.
While we overview the literature in Section~\ref{sec:related}, it is worth noting that most of the studies employ  CT images only. 
However, the integration of complementary information from CT and PET has been shown to improve the accuracy of histological subtype classification~\cite{jacob2022pathological}.
The progression from processing and learning from a single data type to multiple modalities represents a significant shift in how DL models mine and integrate information, with promising results in healthcare~\cite{guarrasi2024systematic}, and it is referred to as multimodal deep learning (MDL). 
MDL techniques can be categorized into three main methods: early, intermediate, and late fusion. 
Early fusion combines features at the raw data level, which can lead to the loss of unique modality-specific traits, while late fusion occurs at the decision level and may overlook deeper interactions between modalities. 
In contrast, intermediate fusion integrates data at the feature extraction stage, offering a more effective combination of modality-specific characteristics.

On these grounds, in this work, we propose a novel approach for intermediate fusion in MDL applied to histological subtype classification in NSCLC.
Our approach automatically fuses knowledge extracted by convolutional neural networks from CT and PET images at different levels of abstraction, enabling gradual integration across multiple layers of the feature hierarchy.
To demonstrate the advantages of our intermediate fusion approach, we evaluated its performance against several benchmarks. 
First, we compared it to unimodal models, which we implemented using the individual branches of our proposed multimodal model, as well as relevant models from the literature. 
Additionally, we compared our approach with other fusion strategies, specifically early and late fusion techniques.
Finally, we benchmarked our model against the only existing study utilizing an intermediate fusion of CT and PET images for histological subtype classification.

\section{Related Work}
\label{sec:related}

Distinguishing between ADC and SQC has become essential in the era of targeted therapies for NSCLC~\cite{travis2015}. 
Current diagnostic methods often rely on invasive procedures that can be both challenging and occasionally inaccurate, highlighting the need for non-invasive techniques to accurately classify NSCLC subtypes: 
for this reason, researchers and practitioners, supported by recent advances in AI, have proposed methods for recognizing histological subtypes extracting the necessary information from CT and/or PET scans e.g.,~\cite{tomassini2022lung,qin2020fine}. 
We can categorize these studies based on the imaging modality used (either CT, PET, or both) or the feature extraction method employed, which primarily consists of hand-crafted radiomic features or learned deep features. 
With respect to the imaging modality, we observe that the majority of these studies employ only CT scans.
Among them, earlier works predominantly represented the information using hand-crafted features~\cite{zhu2018radiomic, linning2019radiomics}, whereas recent approaches have increasingly leveraged deep learning techniques to extract deep features, demonstrating a shift towards more advanced and automated feature extraction methods, as the readers can deepen in the survey~\cite{tomassini2022lung}. 
Conversely, fewer studies focus solely on PET images (e.g,~\cite{sha2019identifying, hyun2019machine}), and none of them apply deep learning techniques for feature extraction. 

It has also been demonstrated that combining the metabolic information from PET scans with the anatomical information from CT scans can enhance the accuracy of classification models~\cite{jacob2022pathological}.
Similarly, earlier studies on multimodal histological subtype classification with PET/CT scans used radiomic features~\cite{yan2020development, koyasu2020usefulness}, while recent works focus on deep networks exploiting end-to-end learning,  integrating feature extraction and classification into a unified framework (\tablename{ \ref{tab:related_works}).
This shift is also motivated by the fact that deep neural networks, and convolutional neural networks (CNNs) in particular,  have the powerful ability to automatically focus on important regions of an image without the need for manual segmentation that, on the contrary, is needed to compute hand-crafted radiomics features. 
Indeed, through the use of convolutional filters, CNNs learn hierarchical features by detecting patterns such as edges, textures, and shapes: these filters highlight the most relevant parts of the image, allowing the network to focus on key areas, as revealed in numerous works by saliency maps~\cite{selvaraju2017grad,jiang2021layercam}. 

\begin{table*}[!htbp]
\caption{\label{tab:related_works}Summary of studies in the literature on multimodal deep learning methods for histological subtype classifications using PET and  CT scans.}
\centering
\begin{tabular}{p{2.9cm}|p{0.8cm}|p{2.6cm}|p{1.6cm}|p{2.1cm}|p{2.4cm}|p{2cm}} 
\toprule
\textbf{Study} & \textbf{Year} & \textbf{Dataset} & \textbf{\# Samples} & \textbf{Input Type} & \textbf{Fusion Method} & \textbf{Model}  \\ 
\midrule
Qin et al.~\cite{qin2020fine} & 2020 & Private & 397 scans & Whole volume & Intermediate & Custom CNN \\\hline
Han et al.~\cite{han2021histologic} & 2021 & Private & 1419 scans & Tumor slice & Early & VGG-16 \\\hline
Jacob \& Menon~\cite{jacob2022pathological} & 2022 & Lung-PET-CT-Dx & 1744 slices & Whole slice & Early & Custom CNN \\\hline
Barbouchi et al.~\cite{barbouchi2023transformer} & 2023 & Lung-PET-CT-Dx & 1160 slices & Whole slice & Early & DETR \\\hline
Zhao et al.~\cite{zhao2024non} & 2024 & Private & 189 scans & Tumor slice  & Early & 7 CNNs \\ \bottomrule
\end{tabular}
\end{table*}

As~\tablename~\ref{tab:related_works} shows, four out of five studies employing deep learning for histological subtype classification use an early fusion strategy to combine CT and PET images, integrating voxel values from both images using an image registration method. Since this combination happens at the raw data level, it may result in the loss of modality-specific characteristics. 
Alternatively, intermediate fusion merges data during feature learning, preserving each modality's unique characteristics while leveraging their complementary information. 
This method is particularly effective for handling complex multimodal biomedical data. 
DL models excel in this approach, as they can capture intricate, nonlinear relationships between modalities, crucial for accurately interpreting their distinct yet complementary information~\cite{guarrasi2024systematic}.
Despite such potential benefits, only one study~\cite{qin2020fine}  has utilized intermediate fusion, indicating that this approach remains relatively unexplored in the context of histological subtype classification.
Indeed, Qin et al.~\cite{qin2020fine} proposed two DenseNet-based CNNs to separately extract features from CT and PET images, followed by a gated multimodal unit to fuse these features, and a fully connected layer to classify histological subtypes. 
It is worth noting that such an approach, which first extracts automatic features using one backbone network per modality, and then merges them at a single point before the classification head, is widely used in the literature~\cite{guarrasi2024systematic}. 
However, this method overlooks the potential benefits of spatial correlations between the modalities, as the features are reduced to a vector at the end of the feature extraction backbones. 
Given that CT and PET images are acquired simultaneously from the same patient in the same position, there is typically a high degree of spatial correlation between them, despite slight misalignments caused by respiratory motion. 
Therefore, in this study, we propose a model that employs a multiple fusion mechanism, where fusion occurs repeatedly throughout the network, starting from the initial layers where the feature maps are still 3D tensors, thus retaining spatial information. 
Additionally, we use these fused features to further guide the network in extracting unimodal features. 
This approach allows us to exploit the complementary information provided by both modalities in greater detail, utilizing voxel-wise rather than scan-wise features.

\section{Materials}
\label{sec:materials}
We combined three datasets — one private and two publicly available — totaling 714 subjects. 
The private dataset consists of 423 patients from the IRCCS Humanitas Research Hospital~\cite{kirienko2018prediction}, selected based on a pathological diagnosis of NSCLC, a baseline [\textsuperscript{18}F]FDG PET/CT scan, and subsequent surgery at the same facility. 
Exclusion criteria included histological types other than ADC or SQC, concomitant cancers, or a history of malignancy within three years before the NSCLC diagnosis, resulting in 312 ADC and 111 SQC cases.

We also included two public datasets. 
The NSCLC Radiogenomics dataset~\cite{radiogenomics} comprises 211 patients from Stanford University and the Palo Alto Veterans Affairs Healthcare System. 
We selected 193 patients from this dataset, with 160 diagnosed with ADC and 33 with SQC, based on the same criteria as the private dataset. 
The Lung-PET-CT-Dx dataset~\cite{lung-pet-ct-dx} contains data from 355 patients who underwent lung biopsy and PET/CT:  as before, we included data only from ADC and SQC cases with both CT and PET images, resulting in 74 ADC and 24 SQC cases.
Ultimately, our final dataset consisted of 546 ADC and 168 SQC cases.

\section{Methods}
\label{sec:method}
We propose an end-to-end deep learning classification pipeline that takes raw CT and PET images as input and outputs the histological subtypes. 
The entire framework is depicted in~\figurename{~\ref{fig:framework}}. 
Our pipeline starts with a pre-processing step shown in panel (a), where the raw scans are prepared for analysis by the network.
The corresponding details are presented in Section~\ref{sec:preprocess}. 
After the pre-processing, the images are fed into a custom 3D multimodal multi-fusion network for classification (panel (b) of ~\figurename~\ref{fig:framework}), with architectural details explained in Section~\ref{sec:architecture}.
It utilizes a multi-stage fusion mechanism, in which fusion occurs repeatedly throughout the network, beginning in the early layers where feature maps still retain 3D spatial information.
This approach allows us to leverage the spatial correlation between modalities.
At each stage of our architecture, features are fused and then redistributed across the unimodal backbones within the same fusion block. 
This strategy enables the network to extract modality-specific features, effectively harnessing the complementary information provided by both imaging modalities in greater detail.

\begin{figure*}[!htbp]
\centering
\includegraphics[width=\textwidth]{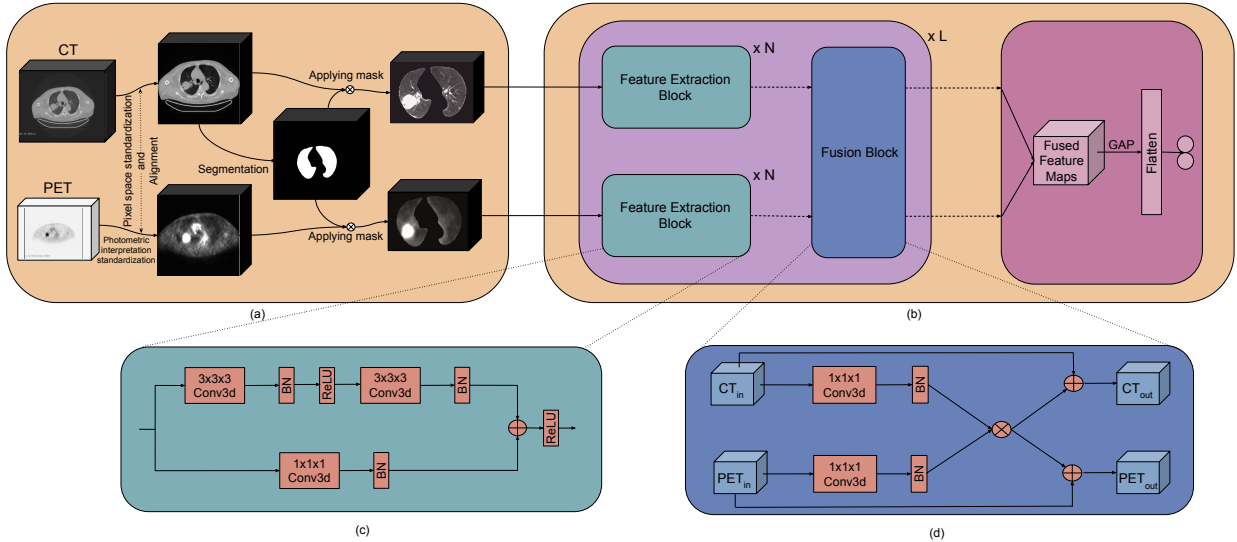}
\caption{Overall framework of the proposed method. (a) Pre-processing. (b) Proposed multimodal convolutional architecture, where $N$ indicates the number of feature extraction blocks in a stage, and $L$ represents the number of stages in the model. (c) Detailed schematic of a feature extraction block. (d) Detailed schematic of a fusion block.}
\label{fig:framework}
\end{figure*}

\subsection{Pre-processing}
\label{sec:preprocess}
CT and PET scans typically encompass large volumes of data, extending beyond the lungs and even outside the body. 
This vast amount of extraneous information presents challenges for deep learning models, as the size of the tumor—being relatively small in comparison to the entire scan—can hinder the networks' ability to effectively learn meaningful features. 
Additionally, variations in scan characteristics due to differences in the imaging machines further complicate the learning process.
To address these issues, we designed a pre-processing pipeline that standardizes the scans and narrows the focus to the region of interest, thereby facilitating feature extraction for the networks. 
First, we standardized the photometric interpretation, addressing variations where some scans utilized higher intensity values for darker regions. 
Second, the intensities of CT scans were converted to Hounsfield Units (HU), while those of PET scans were transformed into Standard Uptake Values (SUV).
Third, we applied linear interpolation to normalize the slice thickness and pixel spacing across all scans; we set the  $xyz$ dimensions to 0.977 mm $\times$  0.977 mm $\times$  3.27 mm, as these were the most prevalent within the dataset. 
Fourth, we aligned the CT and PET scans to ensure consistent origins and endpoints.
Fifth, we used a well-established segmentation algorithm~\cite{hofmanninger2020automatic}   to segment the lungs from the CT images, and the resulting lung masks were applied to both CT and PET scans. 
In the sixth step, we clipped the pixel intensities of the CT scans to the range of [-1024, 1024] and the PET scans to [0, 20], and then we normalized all scans to have voxel values within the range of [0, 1], ensuring uniformity for subsequent processing by deep learning models.

\subsection{Network architecture}
\label{sec:architecture}

We designed a network architecture that aims at extracting and fusing features from both imaging modalities (CT and PET) simultaneously, allowing the modalities to guide each other throughout the feature extraction process. 
To this goal, the overall network is organized in $L$ {\em stages}, represented in violet in panel (b) of \figurename~\ref{fig:framework}, each performing feature extraction and fusion.
Within each stage, we have $N$ feature extraction blocks per modality represented in dark green in the figure and described in Section~\ref{sec:feature_extract}.
After the computation of the automatic features, each stage has one fusion block, shown in blue and presented in Section~\ref{sec:fusion}.
Note that we refer to $L$ stages and $N$ feature extraction blocks because it is possible to experimentally customize this architecture on a validation set.
In this respect, we provide the specific details, including the parameters of the network and the training procedure, in Section~\ref{subsec:NetworkConfig}.
At the end of this modular architecture, we concatenate the fused feature maps from both modalities along the channel dimension. 
This latent space is then passed through the classification head which contains a global average pooling layer that averages the spatial dimensions to generate a feature vector, and a fully connected output layer with two neurons, enabling the model to make final predictions.

\subsubsection{Feature extraction block}
\label{sec:feature_extract}
The feature extraction blocks in our model are responsible for extracting deep features, each utilizing the basic block of well-established 3D ResNet architecture~\cite{tran2018closer}, which has proven effective across a wide range of domains. 
This block consists of a main branch and a residual branch, enabling the construction of deeper networks by mitigating the vanishing gradient problem. 
Consequently, it allows us to increase the number of feature extraction blocks ($N$) and stages ($L$) in our implementation. 
As depicted in panel (c) of \figurename~\ref{fig:framework}, the main branch begins with a $3 \times 3 \times 3 $ convolutional layer, followed by a batch normalization layer and a ReLU activation function. 
This is succeeded by another $3 \times 3 \times 3 $ convolutional layer,  followed again by a batch normalization layer.
In parallel, the residual branch includes a $1 \times 1 \times 1 $ convolutional layer,  followed by batch normalization.
This convolution is necessary to ensure the spatial dimensions align with the main branch, particularly when the main branch reduces the spatial dimensions.
The outputs of the two branches are combined through element-wise summation, and a ReLU activation function is then applied.
As shown in panel (b) of \figurename~\ref{fig:framework}, the output of a block can be passed either to the next feature extraction block (since there could be $N$ block) or to a fusion block.
In the initial block of each stage, the first convolutional layer uses a stride of 2 and produces $2\times C$ feature maps, where 
$C$ represents the number of feature maps in the final layer of the preceding stage. 
This convolutional layer reduces the spatial dimensions of the input, addressing the absence of pooling layers throughout the network. Additionally, it enhances the representational capacity of the extracted features by doubling the number of feature maps. The subsequent convolutional layers within the stage use a stride of 1 and also produce $2\times C$ feature maps, maintaining the increased depth of the feature maps while preserving spatial dimensions.

\subsubsection{Fusion block}
\label{sec:fusion}

Our core idea is that CT and PET images offer complementary insights that can mutually enhance feature extraction and, hence, the information available for each patient. 
To this goal, we design each fusion block to combine data from both modalities by performing element-wise multiplication.
Furthermore, we introduce two residual branches to incorporate the fused features back into the original unimodal data.

As illustrated in panel (d) of~\figurename~\ref{fig:framework}, we denote as $CT_{in}$ and $PET_{in}$ the feature maps corresponding to the outputs of the previous basic blocks, which extract the features from  CT and PET branches, respectively.  
These feature maps are first passed through a $1 \times 1 \times 1 $ convolutional layer with an output feature map size of 1, squeezing the feature maps along the channel dimension and yielding a single feature map for each modality. 
After a  batch normalization step  applied to both modalities, we introduce an element-wise multiplication  between the two feature maps.
The resulting fused feature map is then added to the original input feature maps, $CT_{in}$ and $PET_{in}$ with element-wise summation, producing the output maps $CT_{out}$ and $PET_{out}$. 
We formalize this fusion process by the following equations:
\begin{equation}
\label{eq:fusion_ct}
    CT_{out} = CT_{in} \oplus BN(f_1^1(CT_{in})) \otimes BN(f_1^1(PET_{in}))
\end{equation}
\begin{equation}
\label{eq:fusion_pet}
    PET_{out} = PET_{in} \oplus BN(f_1^1(CT_{in})) \otimes BN(f_1^1(PET_{in}))
\end{equation}
where $BN$ denotes the batch normalization, and $f_k^c$ represents a convolutional layer with a channel size of $c$ and a kernel size of $k \times k \times k$. The symbols $\oplus$ and $\otimes$ are used to indicate element-wise addition and element-wise multiplication, respectively.

\subsection{Network configuration}
\label{subsec:NetworkConfig}

In the previous section, we reported that the network consists of $N$ feature extraction blocks and $L$ stages so that it is possible to identify the best configuration for a given task. 
To this end, we performed a grid search varying both $N$ and $L$ in the range $[1,5]$. We applied stratified 5-fold cross-validation after shuffling the three datasets included in this work, so that training, validation, and test sets account for $60\%$, $20\%$, and $20\%$ of samples, respectively. 
Straightforwardly, the architecture search was conducted on the validation set.
During these experiments we trained the models for 100 epochs, with the learning rate reduced by a factor of 0.1 every 25 epochs,   using the Adam optimizer with class weights and an initial learning rate equal to 0.001. 
We evaluated performance using accuracy, the area under the receiver operating characteristic curve (AUC), and the geometric mean of sensitivity and specificity (Gmean). 
In particular, we consider AUC and Gmean since the a-priori class distribution is imbalanced. 
Indeed, the former focuses on the model's ranking ability, evaluating how well a model differentiates between the two classes regardless of the class distribution.
The latter offers a balanced assessment of the two classes ensuring that the model performs well for both. 
The results of this grid search showed us that the best-performing architecture consists of three stages, with three blocks per stage.
Furthermore, we found that models with only one stage performed the worst, regardless of the number of blocks, suggesting that multiple stages of fusion improve overall performance.
 Regarding the number of feature maps throughout the network, we selected 16 feature maps for the first convolutional layer in the first block of the initial stage. 
Consequently, all convolutional layers within the first stage output 16 feature maps, while the second stage outputs 32, and the third stage outputs 64. 
As a result, the final feature vector, combining information from both modalities, consisted of 128 features.
\begin{figure}[!htbp]
\centering
\includegraphics[width=\columnwidth]{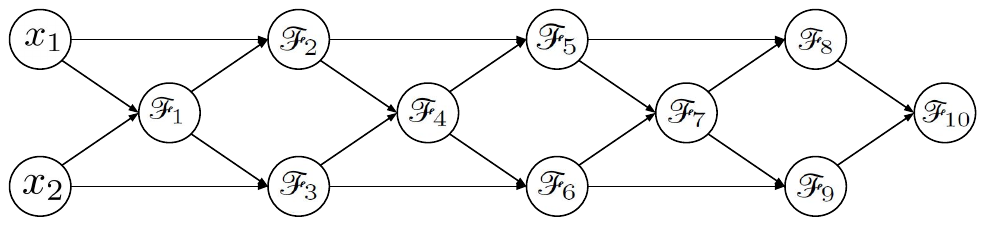}
\caption{Graphical representation of the overall fusion mechanism.}
\label{fig:fusion}
\end{figure}

Given the network configuration found, \figurename{~\ref{fig:fusion}} represents the overall fusion mechanism of our  approach   using the graph representation proposed in~\cite{guarrasi2024systematic}.
The nodes $x_1$ and $x_2$ represent the input modalities, CT and PET, respectively, while each $\Fusion{j}$ denotes a fusion event occurring within the network.
Still, according to~\cite{guarrasi2024systematic}, this process can be formalized by the following equations:
\begin{equation} \label{eq1Fusion}
    \Fusion{3i+1} = \otimes(\Fusion{3i-1}^7,  \Fusion{3i}^7)
\end{equation}
\begin{equation} \label{eq2Fusion}
    \Fusion{3i+2} = \oplus(\Fusion{3i-1}^6, \Fusion{3i+1}^0)
\end{equation}
\begin{equation} \label{eq3Fusion}
    \Fusion{3i+3} = \oplus(\Fusion{3i}^6, \Fusion{3i+1}^0)
\end{equation}
where $\Fusion{j}^l$ represents a fusion operation, except for $\Fusion{-1}$ and $\Fusion{0}$, which represent $x_1$ and $x_2$ in \figurename{~\ref{fig:fusion}}, the inputs CT and PET, respectively. The subscript $j$ represents the fusion number while the superscript $l$ represents the number of trainable layers in which fusion inputs have been processed before the fusion.  
Finally, the symbols $\otimes()$ and $\oplus()$ represent element-wise multiplication and element-wise summation, respectively, and $i$ represents the stage index, which ranges from 0 to 2. 
 Finally, the model is finalized with the following step of fusion:
\begin{equation}
    \Fusion{10} = concat(\Fusion{8}^0, \Fusion{9}^0)
\end{equation}
where $concat()$ indicates a concatenation operation.

\section{Results}
\label{sec:results}
We performed a series of experiments to assess the performance of our proposed model, aiming to compare our multimodal approach with seven competitors. 
They are:
i) four unimodal models that rely exclusively on either CT or PET imaging, 
ii) two alternative fusion strategies, i.e., early and late fusion methods, and 
iii) the only existing study utilizing intermediate fusion of CT and PET images for histological subtype classification~\cite{qin2020fine}.

In case i), we tested four different unimodal models. 
Two, named as {\em  CT Branch} and {\em PET Branch}, use the corresponding branch in our architecture, i.e., each one is a network with 3 stages, each containing 3 blocks, and a classification head at the end but without any fusion block. 
We selected the other two unimodal models from the literature: DetectLC~\cite{fathalla2022detect} and LUCY~\cite{tomassini2023cloud}, chosen because they classify histological subtypes using a 3D approach on CT lung volumes,  similar to our unimodal approach in terms of input structure. 
\begin{table}[ht]
\caption{Average results across 5 folds, presented as mean (standard deviation). 
The highest scores for each metric are highlighted in bold.}
\label{table:results}
\centering
\begin{tabular}{l|l|c|c|c}
\toprule
\multicolumn{2}{c|}{ \textbf{Model} } & \textbf{Accuracy}  & \textbf{AUC} & \textbf{Gmean}  \\ 
\midrule
\multirow{4}{*}{\rotatebox{90}{Unimodal}} &  CT branch & .607 (.168) & .489 (.096) & .305 (.283) \\ 
 & PET branch & .624 (.206) & .465 (.153) & .329 (.109) \\ 
 &  DetectLC~\cite{fathalla2022detect} & .342 (.237) & .499 (.003) & .000 (.000)\\ 
 & LUCY~\cite{tomassini2023cloud} & \textbf{.762 (.008)} & .641 (.061) & .175 (.194) \\ 
\midrule
\multirow{4}{*}{\rotatebox{90}{Multimodal}} & Early fusion & .655 (.164) & .452 (.078) & .224 (.218) \\ 
 & Late fusion & .657 (.109) & .513 (.116) & .342 (.244) \\ 
 & Qin et al.~\cite{qin2020fine} & .539 (.164) & .421 (.073) & .280 (.171)\\ 
 & Our proposal & .724 (.030) & \textbf{.681 (.042)} &\textbf{.646 (.062)} \\ 
\bottomrule
\end{tabular}
\end{table}

Case ii) tests early and late fusion by using the same unimodal architecture as before, i.e., two branches with three stages, each containing three blocks and a classification head at the end of each branch, but without any joint fusion blocks.
To set up the early fusion, we merged the CT and PET images before feeding them into the network using element-wise multiplication, as in our multimodal approach.  
For late fusion, we again used the two separate unimodal branches and then we averaged their output probabilities to make predictions during inference.
Finally, in case iii) we compared our method with the only existing intermediate fusion approach for PET/CT histological subtype classification~\cite{qin2020fine}, which employs a single-fusion block that integrates the modalities after extracting individual feature vectors from each modality using two separate branches, as described in Section~\ref{sec:related}. 
Even though these branches are trained with a shared loss, the effect of each modality on the other remains at a high level of abstraction because the feature fusion occurs only once and before the classification head.
In contrast, we have presented a multi-fusion method, where the fusions occur at various levels of the feature extraction hierarchy, preserving spatial correlations embedded in the  feature maps and allowing for more extensive information sharing between modalities. 

\tablename{~\ref{table:results}}  presents the results attained by such seven competitors and by our method, displaying the average accuracy, AUC, and Gmean scores computed across the five cross-validation runs.
Focusing on the results of the unimodal approaches, we notice that our multimodal method outperforms these competitors in all metrics, except for accuracy in the case of LUCY.
We also observe that the two unimodal competitors drawn from the literature achieve the lowest Gmean scores, suggesting a bias toward one class; in particular,  DetectLC collapses into a single class across all folds. 
Although LUCY demonstrated the highest accuracy and a comparable AUC score, its Gmean ranks as the second-worst: 
this suggests that it struggles to effectively predict the minority class, i.e.,  SQC in our dataset, and LUCY's high accuracy is likely a result of significant bias toward the majority class.
To deepen this analysis we also run the Wilcoxon signed-rank test on the AUC and Gmean scores, as these metrics better represent performance given the data's imbalance.
In all pairwise comparisons for the Gmean score, our approach statistically differs from the unimodal approaches ($p < 0.05$).
The same consideration holds for the AUC score, except when comparing with LUCY $(p = 0.16)$. 
It is also worth noting two unimodal baselines (CT and PET branches) are derived from our network and, hence, their comparison with our approach is equivalent to an ablation test.
This observation, together with the previous ones, supports the consideration that neither modality alone captures the full range of meaningful features necessary for an effective classification.

Let us now turn our attention to the results of multimodal approaches in \tablename{~\ref{table:results}}.
We notice that our approach outperforms the other three in all metrics; the Wilcoxon signed-rank test shows that our performance statistically differs from all competitors for both AUC and Gmean ($p<0.05$), except for Gmean in the case of late fusion where we get $p= 0.0625$, which is close to the significance threshold.
Furthermore, early fusion achieves a lower Gmean score than the unimodal backbones, suggesting that data-level fusion might even harm the classification model. 
While late fusion shows some improvement over early fusion, both methods still fall short of the proposed intermediate fusion approach, which demonstrates that fusion during the feature extraction process performs better than at the data or decision level.
Furthermore, the sharp increase in all metrics compared to~\cite{qin2020fine} demonstrates that our multi-stage voxel-wise fusion approach performs significantly better than a single-stage fusion of extracted features, highlighting the advantage of integrating features at multiple stages to better capture the complementary information between CT and PET modalities. 

\section{Conclusion}
\label{sec:conclusion}

In this work, we have presented a novel multimodal approach for histological subtype classification in NSCLC, utilizing an intermediate fusion method that automatically integrates CT and PET images at various network depths. 
Our experiments show the effectiveness of this approach in comparison to unimodal baselines and other fusion techniques, being also able to handle the challenges posed by dataset imbalance.
By harnessing the complementary information from both imaging modalities, we underscore the value of multimodal fusion in medical image analysis to provide a more comprehensive understanding of tumor characteristics.

Despite these promising results, our work has certain limitations that point to two potential areas for future development. 
First, we aim to refine the proposed multimodal approach by integrating genomics data with imaging features to improve classification accuracy and provide a more comprehensive understanding of tumor biology. 
Second, to enhance the model’s generalizability, we plan to expand the dataset to include additional NSCLC subtypes.

\section*{Acknowledgments}
Fatih Aksu is a PhD student enrolled in the National PhD in Artificial Intelligence, XXXVII cycle, course on Health and life sciences, organized by Università Campus Bio-Medico di Roma.
This work was partially supported by: i)  PNRR MUR project PE0000013-FAIR, ii)  PRIN 2022 MUR 20228MZFAA-AIDA (CUP C53D23003620008), iii) PRIN PNRR 2022 MUR P2022P3CXJ-PICTURE (CUP C53D23009280001), iv) PNRR-MCNT2-2023-12377755.
Resources are provided by the Swedish National Academic Infrastructure for Supercomputing and by National Infrastructure for Computing (SNIC) at Alvis @ C3SE. 

\bibliographystyle{plain}
\bibliography{refs.bib}

\end{document}